\documentclass[%
 aip, apl,
 amsmath,amssymb,
 reprint,%
]{revtex4-1}

\usepackage{graphicx}
\usepackage{dcolumn}
\usepackage{bm}

\usepackage[utf8]{inputenc}
\usepackage[T1]{fontenc}
\usepackage{mathptmx}
\usepackage{etoolbox}
\usepackage{siunitx}
\newcommand{\sicnu}{{College of Physics and Electronic Engineering and Center for Computational Sciences, Sichuan Normal University, Chengdu 610068, China}}

\makeatletter
\def\@email#1#2{%
 \endgroup
 \patchcmd{\titleblock@produce}
  {\frontmatter@RRAPformat}
  {\frontmatter@RRAPformat{\produce@RRAP{*#1\href{mailto:#2}{#2}}}\frontmatter@RRAPformat}
  {}{}
}%
\makeatother
\begin{document}

\preprint{AIP/123-QED}

\title[Sample title]{Cold source field-effect transistor with type-III band-aligned HfS$_2$/WTe$_2$ heterostructure}

\author{Shujin Guo}
\email{kongxianghuaphysics@szu.edu.cn}
\affiliation{\sicnu}

\author{Qing Shi}
\affiliation {Centre for the Physics of Materials and Department of Physics, McGill University, Montreal H3A 2T8, Canada}

\author{Deping Guo}
\affiliation{\sicnu}

\author{Fei Liu}
\email{feiliu@pku.edu.cn}
\affiliation{School of Integrated Circuit, Peking University, Beijing 100871, China}
\affiliation{Beijing Advanced Innovation Center for Integrated Circuits, Beijing 100871, China}

\author{Xianghua Kong}
\email{guosj@sicnu.edu.cn}
\affiliation{College of Physics and Optoelectronic Engineering, Shenzhen University, Shenzhen 518060, China}

\author{Yonghong Zhao}
\affiliation{\sicnu}

\author{Hong Guo}
\affiliation{Centre for the Physics of Materials and Department of Physics, McGill University, Montreal H3A 2T8, Canada}

\date{\today}

\begin{abstract}
The cold source field-effect transistor (CSFET) is promising for reducing power dissipation in integrated circuits by engineering the density of states at the injecting source. Existing CSFET designs utilizing Dirac-source metals or p-Metal-n stacks are challenged by Schottky barriers at the metal-semiconductor interface. In this work, a 2D WTe$_2$/HfS$_2$ heterojunction with type-III band alignment is proposed to be an excellent design of cold source and CSFET. The architecture has a high band-to-band transport mechanism by removing the detrimental Schottky barrier issues. Importantly, the proposed CSFET has the same channel barrier modulation principle as conventional MOSFET to enable a high on-state current. Using first-principles-based quantum transport modeling, we predict a very high $I_{\rm on}$/$I_{\rm off}$ ratio at $\sim$ 10$^{10}$, a low subthreshold swing below the thermal limit for a wide range of gate voltages, reaching as small as 41.3 mV/dec, at low source-drain bias $V_{DS}=0.3$ $\rm V$. These findings establish a design principles for next-generation low-power nanoelectronic switches leveraging 2D van der Waals heterostructures.
\end{abstract}

\maketitle

\textbf{Introduction.} Power dissipation remains a fundamental challenge to continued scaling and integration in modern integrated circuits. A path towards low power consumption is offered by transistors with subthreshold swing (SS) less than the thermodynamic limit of 60 mV/dec. For conventional metal-oxide semiconductor field-effect transistor (MOSFET), the minimum SS is 60 mV/dec at room temperature as a result of the Boltzmann distribution of the injected carriers. Mathematically, the SS is defined as
\begin{equation}
    SS = \frac{\partial V_{\rm G}}{\partial \log I_{\rm DS}},
\end{equation}
where $V_G$ is the gate voltage and $I_{DS}$ is the source-drain current. A smaller SS means smaller $V_G$ for a given current $I_{DS}$ at room temperature, leading to lower power dissipation. By introducing device physics ideas on the switching mechanisms, sub-60 mV/dec SS has been achieved in several FET designs including the tunnel FET (TFET) \cite{appenzeller2004band,szabo2018ab,boucart2007double,cao2015computational,ozccelik2016band,xia2018effects,iordanidou2022two}, the negative capacitance FET (NCFET) \cite{salahuddin2008use,khan2015negative,alam2019critical,jo2016negative,guo2020negative}; and more recently, the cold source FET (CSFET)\cite{liu2018first,liu2018dirac,xu2024realization,logoteta2020cold,wang2021tunneling,zhou2024quantum}.

Specifically, CSFET has been demonstrated using materials such as Dirac metals\cite{qiu2018dirac}, Dirac-like semimetals\cite{liu2020switching}, and engineered semiconductor junctions like pSi-metal-nSi\cite{liu2018first,gan2020design}, which show promising device performance regarding SS. Nevertheless, a significant challenge in these architectures is the formation of a Schottky barrier at the metal-semiconductor interface due to work-function mismatch which degrades carrier injection\cite{louie1976electronic}. Furthermore, Fermi-level pinning caused by metal-induced gap states (MIGS) can exacerbate this issue and impede performance\cite{kobayashi2009fermi,sotthewes2019universal}. To mitigate these interfacial problems of CSFET, here we propose integrating a two-dimensional (2D) van der Waals heterojunction (vdWH) with type-III band alignment as the ``cold'' metal to connect the monolayer components. Type-III band alignment refers to the valence band maximum (VBM) of one layer lying above the conduction band minimum (CBM) of the adjacent layer\cite{ozccelik2016band}. In TFET \cite{cao2015computational,ozccelik2016band,xia2018effects,iordanidou2022two}, type-III alignment facilitates efficient band-to-band tunneling for carrier injection and achieves low-resistance contacts for 2D complementary electronics \cite{yang2023type}. Crucially, the van der Waals interface naturally suppresses dangling bonds and resultant trap states, minimizing Fermi-level pinning\cite{liu2016van}. The efficient tunnel injection and a clean, low-barrier interface should make type-III vdWH an ideal materials platform for realizing a high-performance cold source for CSFET, which we theoretically investigate in this work.

The selection of material is critical for the performance. In this work, WTe$_2$ and HfS$_2$ are selected to construct a type-III CSFET. Both WTe$_2$ and HfS$_2$ are transition-metal dichalcogenides (TMDCs) that feature a small lattice mismatch and have been extensively studied in electronic and photoelectric devices, making the CSFET with WTe$_2$/HfS$_2$ vdWH of high practical feasibility~\cite{lei2019broken,fei2017edge,xu2015ultrasensitive,lv2019band}. Moreover, HfS$_2$ stands out for its high calculated room-temperature mobility of ~1800~cm$^2 {\rm V}^{-1} {\rm s}^{-1}$, which significantly exceeds that of other TMDCs such as MoS$_2$, leading to high on-state currents as the channel material of FETs~\cite{xu2015ultrasensitive}. We consider a 2D CSFET model by laterally integrating monolayer WTe$_2$, WTe$_2$/HfS$_2$ vdWH and monolayer HfS$_2$. This model leverages type-III alignment to achieve two critical functions simultaneously: (1) energy-filtering ``cold'' carrier injection for steep switching, and (2) an inherent Ohmic contact to eliminate detrimental Schottky barriers. The current-voltage ($I_{DS}-V_G$) characteristics of CSFET show high on-state current, large on/off ratio, and a low SS, demonstrating good FET performance.

\textbf{Properties of cold source.} We first perform an analysis for $\text{WTe}_2$, $\text{HfS}_2$, and $\text{WTe}_2/\text{HfS}_2$ heterojunction to confirm the type-III band alignment and understand its formation mechanism, using density functional theory (DFT) as implemented in the VASP package where the electron-ion potential is described by the projector augmented-wave method, the exchange correlation interaction is described by the Perdew-Burke-Ernzerhof (PBE) generalized gradient approximation (GGA), and Opt-B86 van der Waals correction is included to account for the vdw interactions\cite{kresse1996efficiency,klimevs2010chemical,klimevs2011van}. The Brillouin zone is sampled by a $7\times7\times1$ kpoint mesh and the cutoff energy is set at 500 eV. The atomic structures are fully relaxed until the force on each atom is less than 0.01 eV/\AA, and the convergence criterion of total energy is set as $10^{-5}$ eV.

\begin{figure*}
\includegraphics[width=0.7\linewidth]{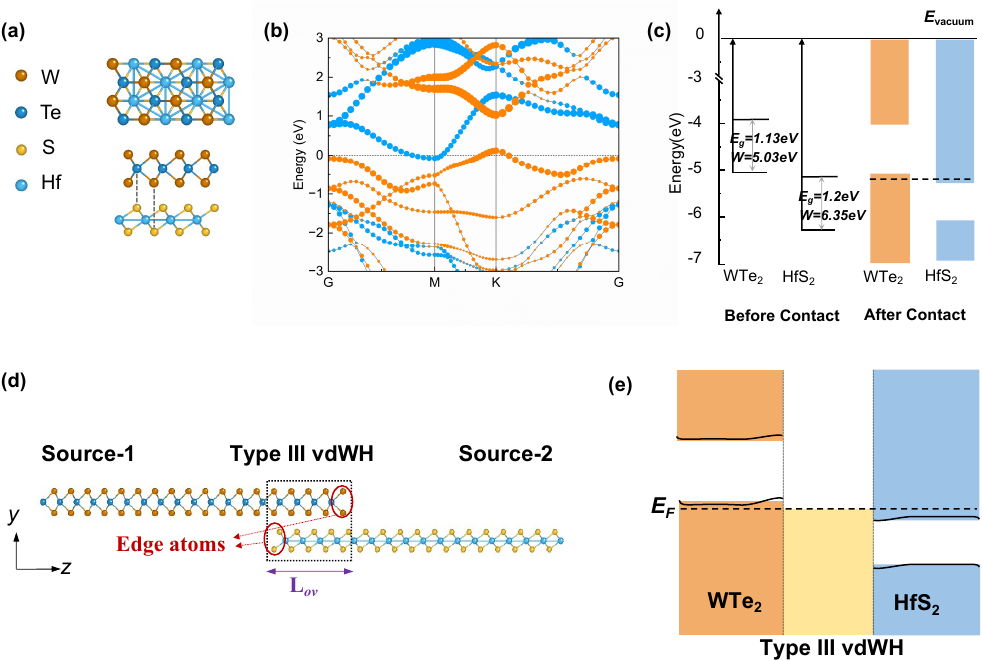}
\caption{\label{fig1} (a) The crystal of WTe$_2$/HfS$_2$ vdWH from the top and side view. WTe$_2$/HfS$_2$ vdWH stacks by fcc-I stacking pattern, in which the W atom is located above the S atom, Te atom aligns with the other S atom. (b) The band structure of WTe$_2$/HfS$_2$ vdWH calculated with PBE functional. (c) The band alignment of WTe$_2$/HfS$_2$ vdWH before and after contact. (d) and (e) are the crystal structure and the band alignment of the cold source. $L_{ov}$ is the overlap length of the vdWH. Edge atoms refer to the most right atoms and the most left atoms in the overlapped region.}
\end{figure*}

Fig.~\ref{fig1}(a) illustrates the top and side views of the relaxed atomic structure of WTe$_2$/HfS$_2$ vdWH. This fcc-I stacking pattern has the lowest formation energy \cite{lei2019broken}. Together with the band structure of WTe$_2$/HfS$_2$ vdWH shown in Fig.~\ref{fig1}(b), the type-III band alignment is clearly observed. More information of the electronic structure of the WTe$_2$ and HfS$_2$ monolayers and related discussions are shown in Fig. S1 of the Supplemental Information (SI), which agrees well with existing literature \cite{lei2019broken}. In Fig.~\ref{fig1}(c), we show the formation of type-III WTe$_2$/HfS$_2$ vdWH. Before contacting each other, the work functions of 2D WTe$_2$ and HfS$_2$ are 5.03 and 6.35 eV, respectively and therefore, upon forming the heterostructure, charges transfer from WTe$_2$ to HfS$_2$. Consequently, the Fermi levels of 2D HfS$_2$ and WTe$_2$ shift upward and downward, respectively, until they align. The resultant work function of vdWH is 5.27 eV which is, as expected, in between those of 2D HfS$_2$ and 2D WTe$_2$. This character of the vdWH lays the foundation to achieve a low-resistance bridge between WTe$_2$ and HfS$_2$ on either side and suggests that it should be advantageous to form a cold source by laterally integrating WTe$_2$, WTe$_2$/HfS$_2$ vdWH and HfS$_2$. Therefore, in the following we investigate the cold source model that consists of three parts: Source-1 (WTe$_2$), vdWH, and Source-2 (HfS$_2$), shown in Fig.~\ref{fig1}(d). Since the differences in work functions between the three parts are small, the Schottky barriers at the interfaces are low. The corresponding band alignment of the cold source model is shown in Fig.~\ref{fig1}(e).

The working principle of the cold source model is as follows. Firstly, the band gap of WTe$_2$ (1.13 eV) is sufficiently high to filter out ``hot'' carriers having high energies, i.e. the band gap of the Source-1 part of the cold source [see Fig.~\ref{fig1}(e)] cuts off the high energy tail of the Fermi-Dirac distribution of the injected carriers. Secondly, the vdWH part smoothly bridges the Source-1 and Source-2 parts with essentially no Schottky barriers at the two interfaces. Thirdly, the type-III band alignment of the Source-2 part (HfS$_2$) with the Source-1 part (WTe$_2$), shown in Fig.~\ref{fig1}(e), completes the cold source, namely electrons tunnel band-to-band directly from the VBM of WTe$_2$ to the CBM of HfS$_2$. Finally, the cold source will be connected to the FET channel at the right of the Source-2 to form the eventual CSFET (see below).

We now proceed to calculate the transport current through the cold source model. The transport properties are calculated with Nanodcal quantum transport package \cite{taylor2001ab,taylor2001} where DFT is carried out within the non-equilibrium Green's function formalism (NEGF-DFT). Monkhorst-Pack k-mesh of 9$\times$1$\times$1 is used. The doping of WTe$_2$ and HfS$_2$ is achieved by the virtual crystal approximation (VCA). The current of the device is calculated with the Landauer-B$\mathrm{\ddot{u}}$ttiker equation,
\begin{equation}
    I = \frac{2q}{h}\int T(E)M(E)[f(E-E_F(S))-f(E-E_F(D))]dE ,
\end{equation}
where $T(E)$ is the transmission probability at energy $E$, $M(E)$ is the number of transport modes, $f(E)$ is the Fermi-Dirac distribution, $E_F(S)$ and $E_F(D)$ are the Fermi levels of source and drain, respectively. The electrode temperature is set to 300 K.

\begin{figure*}
\includegraphics[width=0.8\linewidth]{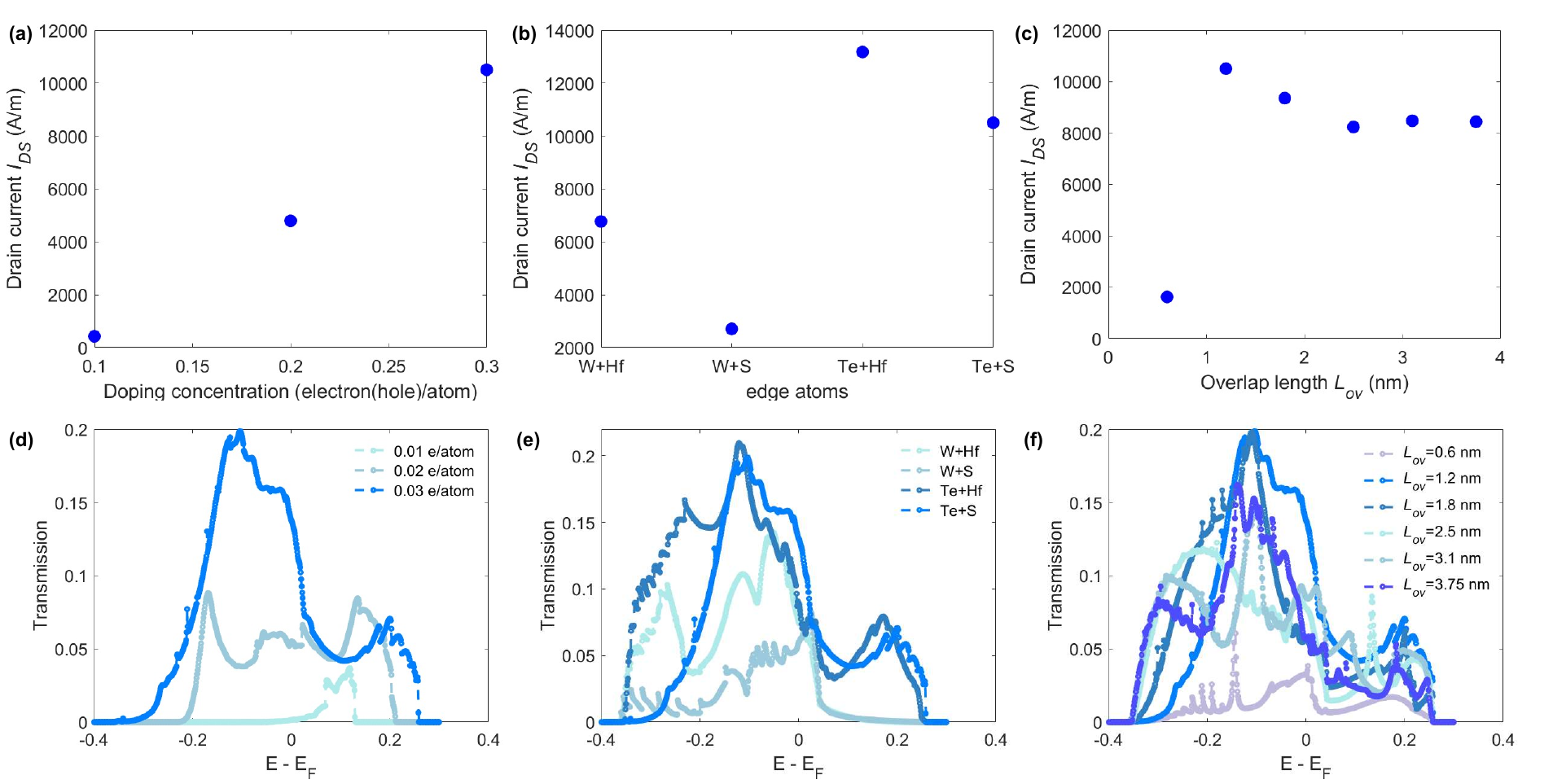}
\caption{\label{fig2} The on-state current of cold sources with (a) different doping concentration, (b) different edge atoms and (c) varied overlap length. (d),(e),(f) are the corresponding transmission spectrum of (a),(b) and (c). The current is calculated by integrating the transmission spectrum between the tunneling window.}
\end{figure*}

\begin{figure*}
\includegraphics[width=\linewidth]{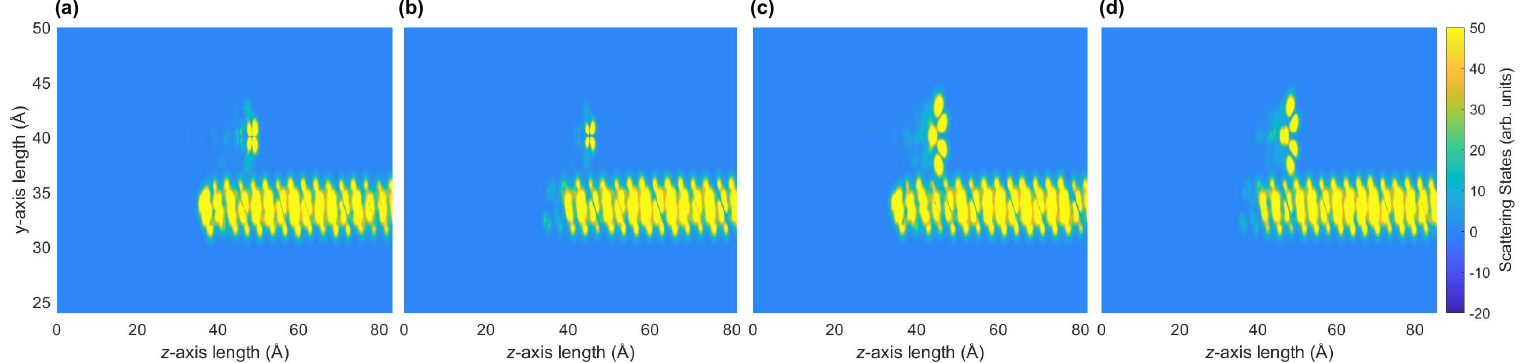}
\caption{\label{fig3} The probability density of scattering states of cold sources with different edge atoms: (a) W+Hf, (b) W+S, (c) Te+Hf, (d) Te+S.}
\end{figure*}

\begin{figure*}
\includegraphics[width=0.6\linewidth]{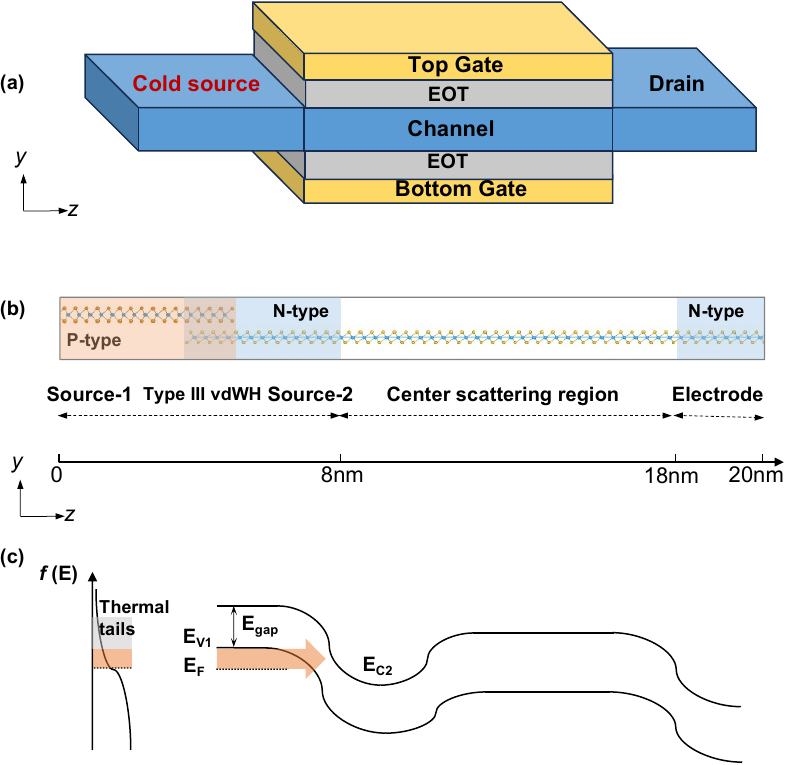}
\caption{\label{fig4} (a) The schematic diagram of the CSFET with double gates. (b) The atomic structure of the entire CSFET, in which the cold source consists of P-type WTe$_2$, WTe$_2$/HfS$_2$ vdWH and N-type HfS$_2$. (c) The band alignment of the whole CSFET.}
\end{figure*}

\begin{figure*}
\includegraphics[width=0.7\linewidth]{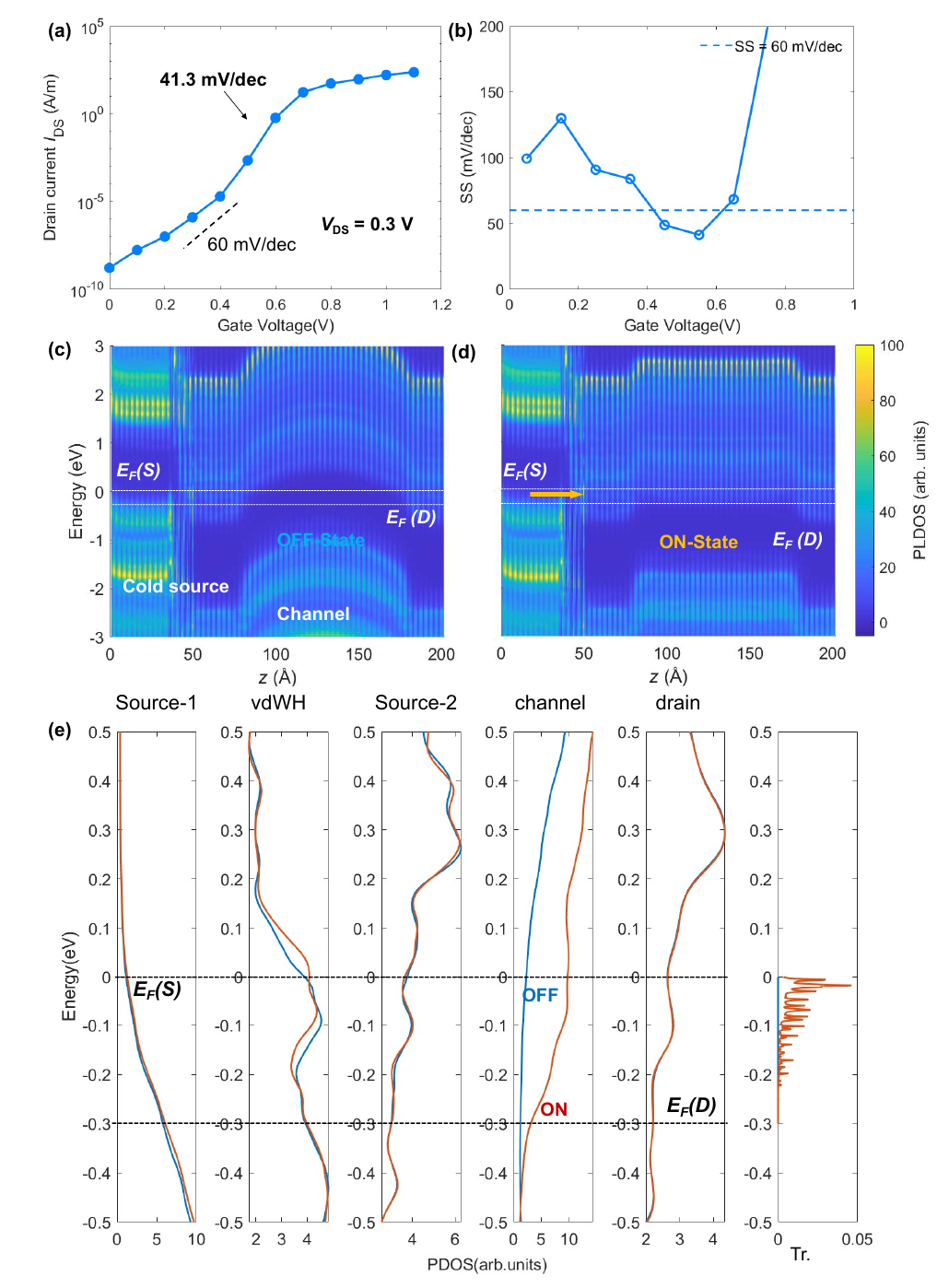}
\caption{\label{fig5} The (a) $I_{DS}-V_G$ curve and (b) corresponding SS of CSFET with $V_{DS}$ = 0.3 V. (c) and (d) are the local density of states (LDOS) of CSFET at off-state ($V_G$ = 0 V) and on-state ($V_G$ = 0.7 V). (e) The projected local density of states (PLDOS) of Source-1, vdWH, Source-2, channel and drain, and transmission spectrum at off-state and on-state. Tr. is the abbreviation of transmission.}
\end{figure*}

The energy range from the WTe$_2$ VBM to the HfS$_2$ CBM defines the transport window in the cold source [see Fig.~\ref{fig1}(e)] whose width can be modulated to an optimal value to achieve a high on-state current while preserving the carrier energy filtering effect of the cold source. Different doping concentrations can be applied to change the band alignment. As shown in Fig.~\ref{fig2}(a), the current increases with increasing doping concentration. Fig.~\ref{fig2}(d) displays the corresponding transmission spectrum for several transport windows, 0.2 eV, 0.46 eV and 0.62 eV, respectively. A higher doping concentration leads to a wider transport window which increases $I_{DS}$. The band alignment of WTe$_2$/HfS$_2$ heterojunction can also be tuned by strain and external electric fields. Both in-plane tension strain and negative electric field expand the VBM and CBM overlap region, leading to a wider transport window\cite{lei2019broken,xia2018effects}. Additionally, the type of edge atoms plays a role in transport properties\cite{logoteta2020cold,balaji2020mos2,cao2015computational}. Cold sources with four edge components are investigated: W+Hf, W+S, Te+Hf and Te+S, with atomic structures shown in Fig. S2. Apparently, devices ending with Te exhibit higher current than those ending with W, while the Hf edges outperform S edges. Among all, the Te+Hf combination yields the highest $I_{DS}$. The transmission spectrum confirms a different contribution of the edge atoms.
 
The probability density of the scattering states in real space [Figs.~\ref{fig3}(a)-(d)] show bright spots at the edge of WTe$_2$ layer, indicating that $I_{DS}$ mainly goes through the edge atoms, regardless of the atom type. Brighter edges appear in Te-terminated WTe$_2$ [Figs.~\ref{fig3}(c), (d)], suggesting higher transport probability through Te. In the HfS$_2$ layer, bright edge spots emerge only when terminated with Hf, aligning with a band-to-band picture: electrons traverse from the VBM of WTe$_2$ (dominated by Te) to the CBM of HfS$_2$ (dominated by Hf)\cite{lv2019band}. Thus, the Te+Hf edge structure maximizes current. Considering the stability of the vdWH edge atoms, in the following we adopt the most stable edge structure, which is the one that ends with Te and S\cite{zhou2017mos2,da2019edge}.

To assess how the overlap length ($L_{ov}$) in the cold source affects the transport property, currents with $L_{ov}$ = 0.63, 1.25, 1.88 and 2.5, 3.13, and 3.75 nm are calculated. The lengths of Source-1 and Source-2 parts are set to the same value. The cold sources' atomic configurations are shown in Fig.~S2(b). As shown in Fig.~\ref{fig2}(c), the current first increases with $L_{ov}$ followed by decreasing to a saturated value. The saturation is the result of the corresponding transmission spectrum [Fig.~\ref{fig2}(e)] which shows wider but lower peaks with increasing $L_{ov}$. When electrons traverse from the upper layer to the lower layer in vdWH, the scattering states diagram in Fig.~S3 shows that transport is mainly through the edge of the WTe$_2$ layer for different $L_{ov}$. In the lower HfS$_2$ layer, more conduction paths appeared in the overlap region with longer $L_{ov}$ since the contact area expanded, leading to higher transport probability. In the vdWH part, there exists a built-in electric field between the layers due to the work function difference\cite{lei2019broken,iordanidou2022two}. This built-in field hinders the lateral movement of the electrons\cite{qin2022van}. Hence, although the contact area is larger for longer $L_{ov}$, the current eventually reaches a saturated value due to the compensating effect of the built-in field.

\textbf{Type III CSFET.} Having determined the properties of cold source, we now proceed to construct the entire type III CSFET shown in Fig.~\ref{fig1}(d) by extending the HfS$_2$ layer to the channel and drain. A schematic diagram and the corresponding atomic structure of the CSFET are shown in Figs.~\ref{fig4}(a,b). As seen, the overlap region (cold source) is kept out of the gate. Additionally, HfO$_2$ is selected as the insulating material inserted between the gate and the channel to improve the gate control and optimize channel carrier modulation. The length of the cold source, channel, and drain are set to 8~nm, 10~nm, and 2~nm, respectively. As discussed above, to reduce contact resistance and achieve a sufficiently wide transport window, WTe$_2$ is p-doped at a concentration of 0.03 holes/atom and HfS$_2$ is n-doped at a concentration of 0.03 electron/atom via the VCA method. The resulting cold source comprises three components: a p-type Source-1 (WTe$_2$), a type-III vdWH, and an n-type Source-2 (HfS$_2$). The channel remains intrinsic and the drain is n-doped, forming an n-type FET with the band alignment schematically shown in Fig.~\ref{fig4}(c).

In this CSFET, carriers tunnel from the valence band ($E_{V1}$) of Source-1 to the conduction band ($E_{C2}$) of Source-2, as indicated by the red arrow in Fig.~\ref{fig4}(c). In the channel, the carriers undergo thermionic injection when the channel barrier is lowered by the gate bias. Unlike a TFET that turns off the current by changing the type-III band alignment to type-II, the switching mechanism of this CSFET is analogous to that of a conventional n-type MOSFET. This design keeps the injection window unaffected by gate modulation which is crucial for maintaining a steep SS across a wide ranges of gate voltage $V_{\rm G}$\cite{smith2011broken}.

The room temperature transport characteristics of the entire CSFET are shown in Figs.~\ref{fig5}(a). The source-drain bias ($V_{DS}$) is fixed to 0.3 V, while the off-state and on-state of the device are defined at $V_G$ = 0 V and $V_G$ = 0.7 V, respectively. Notably, $I_{\rm on}$ can be as large as 2.3$\times$10$^2$ A/m, while $I_{\rm on}$/$I_{\rm off}$ exceeds 10$^{10}$ simultaneously, which shows better switching properties than cold source constructed with p-Si/metal/n-Si structure\cite{liu2018first}. The $I_{\rm on}$ is also somewhat higher than the largest on-state current which is about 2.1$\times$10$^2$ A/m, achieved in 2D graphene Dirac Source FET\cite{liu2018dirac}. Significantly, the SS shown in Fig.~\ref{fig5}(b) is less than 60 mV/dec from 0.4 V to 0.6 V as the ``hot'' thermal carriers are efficiently eliminated by the cold source in the subthreshold regime. The minimum SS is as low as 41.3 mV/dec.

In order to better understand the working mechanism of the device, Fig.~\ref{fig5}(c) shows the local density of states (LDOS) of the entire device at the off-state and on-state, respectively. As seen, the cold source maintains a type-III band alignment that is not affected by gate modulation during the switching process. The density of states are continuous at the vdWH which indicates a metallic behavior. There is no observable Schottky barriers between Source-1, vdWH and Source-2, demonstrating ohmic contact by the vdWH bridge. The white dashed lines refer to the chemical potential of the cold source and the drain, respectively, which defines the bias window. It can be seen that the initial barrier height of the channel is about 0.43 eV. When the channel barrier is higher than the VBM of WTe$_2$, the thermal carriers will not change. When the barrier is reduced below the VBM of WTe$_2$, the current suddenly increases, leading to low SS, and the device is turned on.

To further understand the device principle, the projected LDOS (PLDOS) of Source-1, vdWH, Source-2, channel, and drain are plotted at different gate voltages, shown in Fig. 5(d). The DOS of Source-1 decays with energy above the Fermi level, which is responsible for the reduction of thermal carriers. With increasing gate voltage, the DOS of channel material in the bias window increases as a result of gate modulation, leading to the appearance of the high transmission peak under an efficient gate modulation.

In conclusion, we have designed and modeled a high-performance cold-source FET based on a type-III two-dimensional WTe$_2$/HfS$_2$ van der Waals heterostructure. The device architecture successfully fulfills the dual purpose of a cold source: it effectively filters out Boltzmann-distributed high energy hot carriers to achieve a remarkably low off-state current ($I{\rm _{off}}$ = 1.6$\times$10$^{-9}$ A/m), while the broken-gap alignment ensures efficient band-to-band transport to deliver a high on-state current ($I{\rm _{on}}$ = 2.3$\times$10$^2$ A/m). This combination results in an exceptional device performance, characterized by a high on/off current ratio exceeding $10^{10}$ and a steep subthreshold swing (SS) for a relatively wide range of gate potentials, with the lowest SS at 41.3 mV/dec. Additionally, the effect of the vdWH configuration on CSFET transport property is investigated. We find that $I{\rm _{on}}$ can be optimized with a wider transport window and edge atoms of Te and Hf, while $L_{ov}$ has a limited effect. Significantly, the device can be fabricated directly by stacking two-dimensional material layers, eliminating the need for a metal interlayer and thus avoiding the associated Fermi-level pinning and Schottky barrier issues that plague conventional metal-semiconductor contacts. This work demonstrates the significant potential of type-III vdWHs in designing next-generation, low-power switching devices.

\appendix

\section{Acknowledgments}

S.J.G. thanks Chen Hu, Mingyan Chen, and Yibin Hu for fruitful discussions on the device physics and the NanoDCAL transport package. S.J.G. and H.G. thank FRQNT of Quebec for partial support and gratefully acknowledge the Digital Research Alliance of Canada for computational allocations where this work is partially done. X.H.K. gratefully acknowledges the financial support from the Shenzhen Science and Technology Innovation Commission under the Outstanding Youth Project (Grant No. RCYX20231211090126026), the National Natural Science Foundation of China (Grants No. 12474173, 52461160327), Department of Science and Technology of Guangdong Province (Grants No. 2021QN02L820). We also acknowledge HZWTECH for providing computation facilities.

\end{document}


\preprint{AIP/123-QED}

\title[]{Supporting Information for ``Cold source field-effect transistor with type-III band-aligned HfS$_2$/WTe$_2$ heterostructure''}

\author{Shujin Guo}
 \affiliation {College of Physics and Electronic Engineering and Center for Computational Sciences, Sichuan Normal University, Chengdu 610068, China}
\email{kongxianghuaphysics@szu.edu.cn}

\author{Shi Qing}
 \affiliation {Department of Physics, McGill University, Montreal H3A 2T8, Canada}
 
\author{Deping Guo}
\affiliation {College of Physics and Electronic Engineering and Center for Computational Sciences, Sichuan Normal University, Chengdu 610068, China}
\author{Fei Liu}%
\email{feiliu@pku.edu.cn}
\affiliation{School of Integrated Circuit, Peking University, Beijing 100871, China}
\affiliation{Beijing Advanced Innovation Center for Integrated Circuits, Beijing 100871, China}

\author{Xianghua Kong}
\email{guosj@sicnu.edu.cn}

\affiliation{%
College of Physics and Optoelectronic Engineering, Shenzhen University, Shenzhen 518060, China.}%

\author{Yonghong Zhao}
\affiliation {College of Physics and Electronic Engineering and Center for Computational Sciences, Sichuan Normal University, Chengdu 610068, China}

\author{Hong Guo}
\affiliation{%
Department of Physics, McGill University, Montreal H3A 2T8, Canada}

\date{\today}

\maketitle
\onecolumngrid

\begin{figure*}
\includegraphics[width=0.6\linewidth]{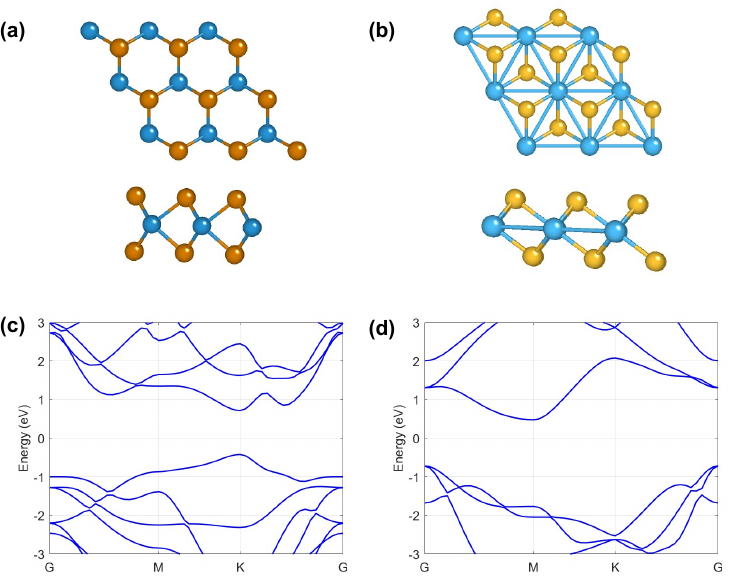}
\caption{\label{fig:s1} The atomic structure and band structure of 2D (a)(c) WTe$_2$ and (b)(d) HfS$_2$ respectively. }
\end{figure*}  
Fig. S1 (a) and (b) display the atomic structure diagram of 2d WTe$_2$ and HfS$_2$ respectively. As can be seen, HfS$_2$ adopts T phase, while WTe$_2$ features H phase. The relaxed lattice constant of WTe$_2$ and HfS$_2$ is 3.64 and 3.55 \AA, respectively. The lattice mismatch between the two layers is 2$\%$. 
Fig. S1 (c) and (d) depict the bandstructure of WTe$_2$ and HfS$_2$ respectively. As is seen, WTe$_2$ has a direct band gap of 1.13 eV, while HfS$_2$ has indirect band gap of 1.2 eV. 

\begin{figure*}
\includegraphics[width=0.8\linewidth]{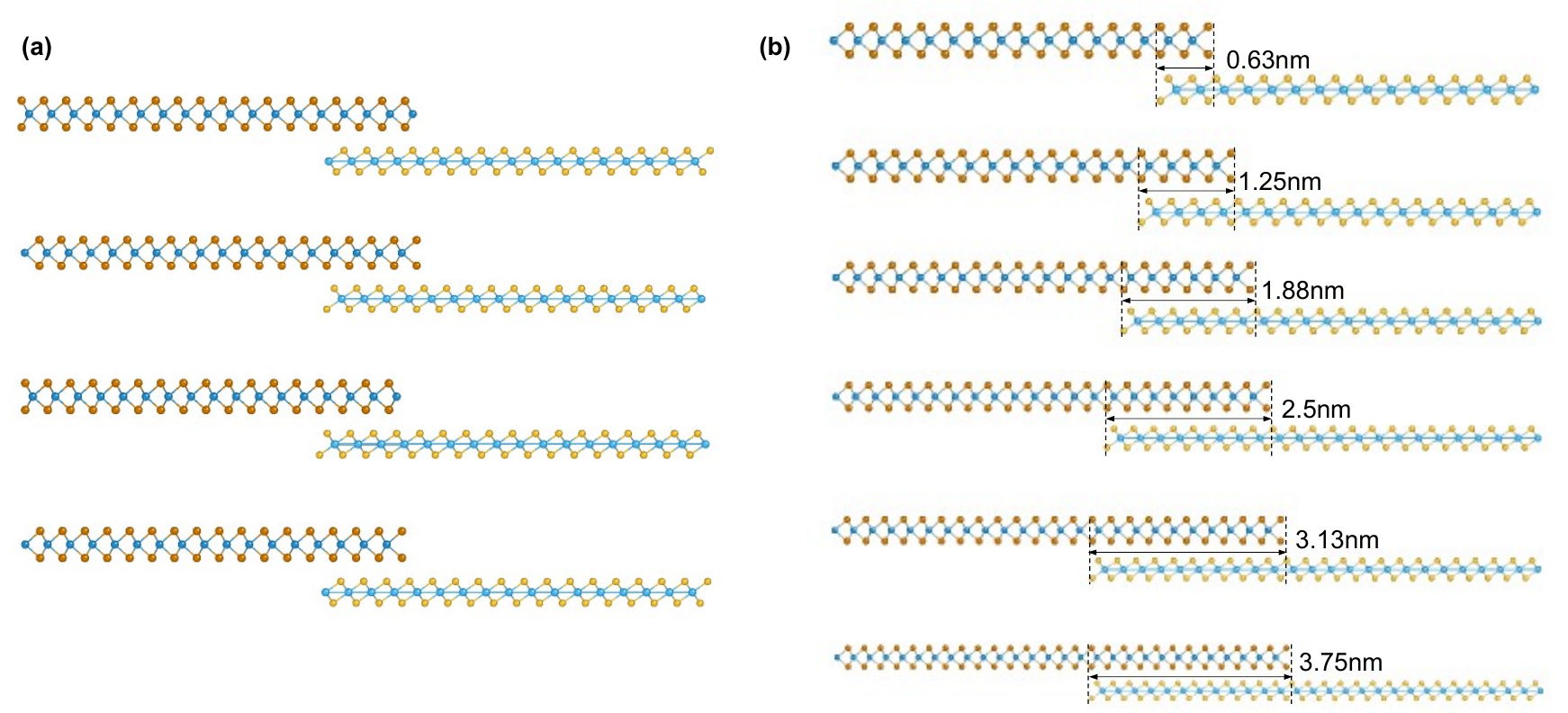}
\caption{\label{fig:s2} The atomic structure diagram of cold source with (a) different edge atoms  and (b) different overlap length. }
\end{figure*}

\begin{figure*}
\includegraphics[width=0.9\linewidth]{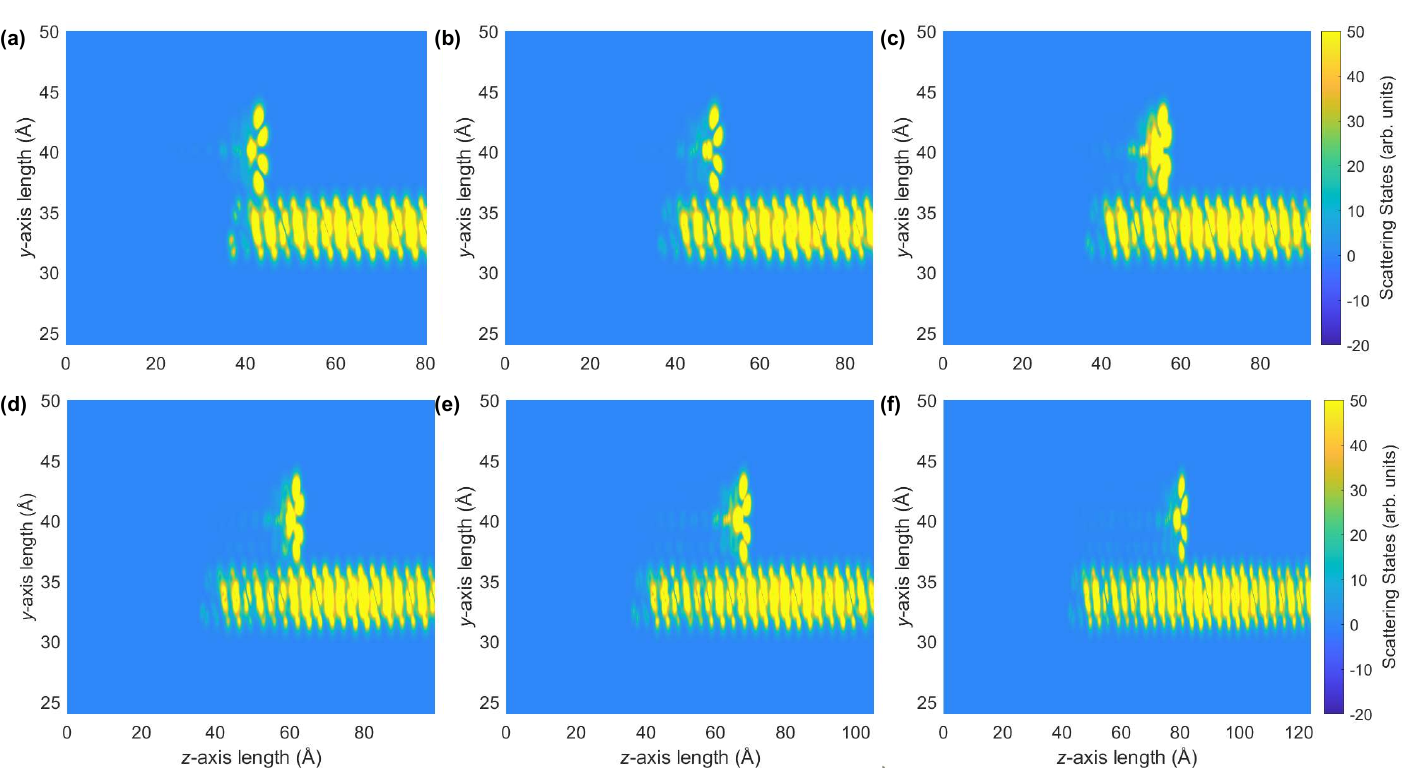}
\caption{\label{fig:s3} 
The probability density of scattering states of cold sources with (a)-(f) $L_{ov}$ = 0.63 nm, 1.25 nm, 1.88 nm, 2.5 nm, 3.13 nm and 3.75 nm respectively.}
\end{figure*}